\documentclass[sn-mathphys]{sn-jnl}

\jyear{2023}%

\theoremstyle{thmstyleone}%
\usepackage{amsmath,bm}
\usepackage{lineno}
\theoremstyle{thmstyletwo}%

\theoremstyle{thmstylethree}%

\makeatother
\def\B1g{B\textsubscript{1g}}

\def\CoF2{CoF$_2$}
\begin{document}

\title[Impulsive Fermi magnon-phonon resonance in antiferromagnetic CoF\textsubscript{2}]{Impulsive Fermi magnon-phonon resonance in antiferromagnetic CoF\textsubscript{2}}

\author*[1]{\fnm{Thomas W.J.} \sur{Metzger}}
\email{thomas.metzger@ru.nl}

\author[1]{\fnm{Kirill A.} \sur{Grishunin}}

 \author[2]{\fnm{Chris} \sur{Reinhoffer}}

\author[3]{\fnm{Roman M.} \sur{Dubrovin}}

\author[4]{\fnm{Atiqa} \sur{Arshad}} 

\author[4]{\fnm{Igor} \sur{Ilyakov}}

\author[4]{\fnm{Thales V.A.G.} \sur{de Oliveira}}

\author[4]{\fnm{Alexey} \sur{Ponomaryov}}

\author[4]{\fnm{Jan-Christoph} \sur{Deinert}}

\author[4]{\fnm{Sergey} \sur{Kovalev}}

\author[3]{\fnm{Roman V.} \sur{Pisarev}}

\author[1]{\fnm{Mikhail I.} \sur{Katsnelson}}

\author[1]{\fnm{Boris A.} \sur{Ivanov}}

\author[2]{\fnm{Paul H. M.} \sur{van Loosdrecht}}

\author[1]{\fnm{Alexey V.} \sur{Kimel}}

\author*[2]{\fnm{Evgeny A.} \sur{Mashkovich}}
\email{mashkovich@ph2.uni-koeln.de}

\affil[1]{\orgdiv{Institute for Molecules and Materials}, \orgname{Radboud University}, \orgaddress{\street{Heyendaalseweg 135}, \city{Nijmegen}, \postcode{6525 AJ}, \country{The Netherlands}}}

\affil[2]{\orgdiv{Institute of Physics II}, \orgname{University of Cologne}, \orgaddress{\street{Zuelpicher Straße 77}, \city{Cologne}, \postcode{10587}, \country{Germany}}}

\affil[3]{\orgdiv{Ioffe Institute}, \orgname{Russian Academy of Sciences}, \orgaddress{\city{St.\,Petersburg}, \postcode{194021}, \country{Russia}}}

\affil[4]{\orgdiv{Institute of Radiation Physics}, \orgname{Helmholtz-Zentrum Dresden-Rossendorf}, \orgaddress{\street{Bautzner Landstraße 400}, \city{Dresden}, \postcode{01328}, \country{Germany}}}
\abstract{
Understanding spin-lattice interactions in antiferromagnets is one of the most fundamental issues at the core of the recently emerging and booming fields of antiferromagnetic spintronics and magnonics. 
Recently, coherent nonlinear spin-lattice coupling was discovered in an antiferromagnet which opened the possibility to control the nonlinear coupling strength and thus showing a novel pathway to coherently control magnon-phonon dynamics.
Here, utilizing intense narrow band terahertz (THz) pulses and tunable magnetic fields up to 7 T, we experimentally realize the conditions of the Fermi magnon-phonon resonance in antiferromagnetic \CoF2. These conditions imply that both the spin and the lattice anharmonicities harvest energy transfer between the subsystems, if the magnon eigenfrequency $f\textsubscript{m}$ is twice lower than the frequency of the phonon $2f\textsubscript{m}=f\textsubscript{ph}$. Performing THz pump-infrared probe spectroscopy in conjunction with simulations, we explore the coupled magnon-phonon dynamics in the vicinity of the Fermi-resonance and reveal the corresponding fingerprints of an impulsive THz-induced response. This study focuses on the role of nonlinearity in spin-lattice interactions, providing insights into the control of coherent magnon-phonon energy exchange.

}




\maketitle
    
\section{Introduction}\label{sec1}
The lattice dynamics of crystals involve interdependent periodic movements of individual atoms, rendering them inherently complex. Nevertheless, the modern theory of condensed matter has successfully managed to describe the dynamics in terms of linear superposition of mutually independent phononic modes comprising acoustic phonons and optical phonons. However, if the amplitude of the lattice vibrations is large, this approximation fails~\cite{maehrlein_terahertz_2017,forst_nonlinear_2011} and the lattice dynamics enter a poorly explored regime, in which a single phonon can be excited by multiple photons and new channels of energy transfer between otherwise non-interacting phononic modes open up. This regime became the main focus of the establishing field of nonlinear phononics~\cite{disa_engineering_2021}. Lately, such a nonlinear regime was also found for magnonic systems~\cite{zhang_three-wave_2023,blank_empowering_2023}. Alternatively one can enhance nonlinearity not by increasing the modulation amplitude but instead by realizing a coupling of eigenstates reaching the nonlinear resonance matching conditions. Surprisingly, this route is barely explored especially under the consideration of accounting for a coupling between different subsystems. 

\begin{figure}
        \centering
        \includegraphics[width=\textwidth]{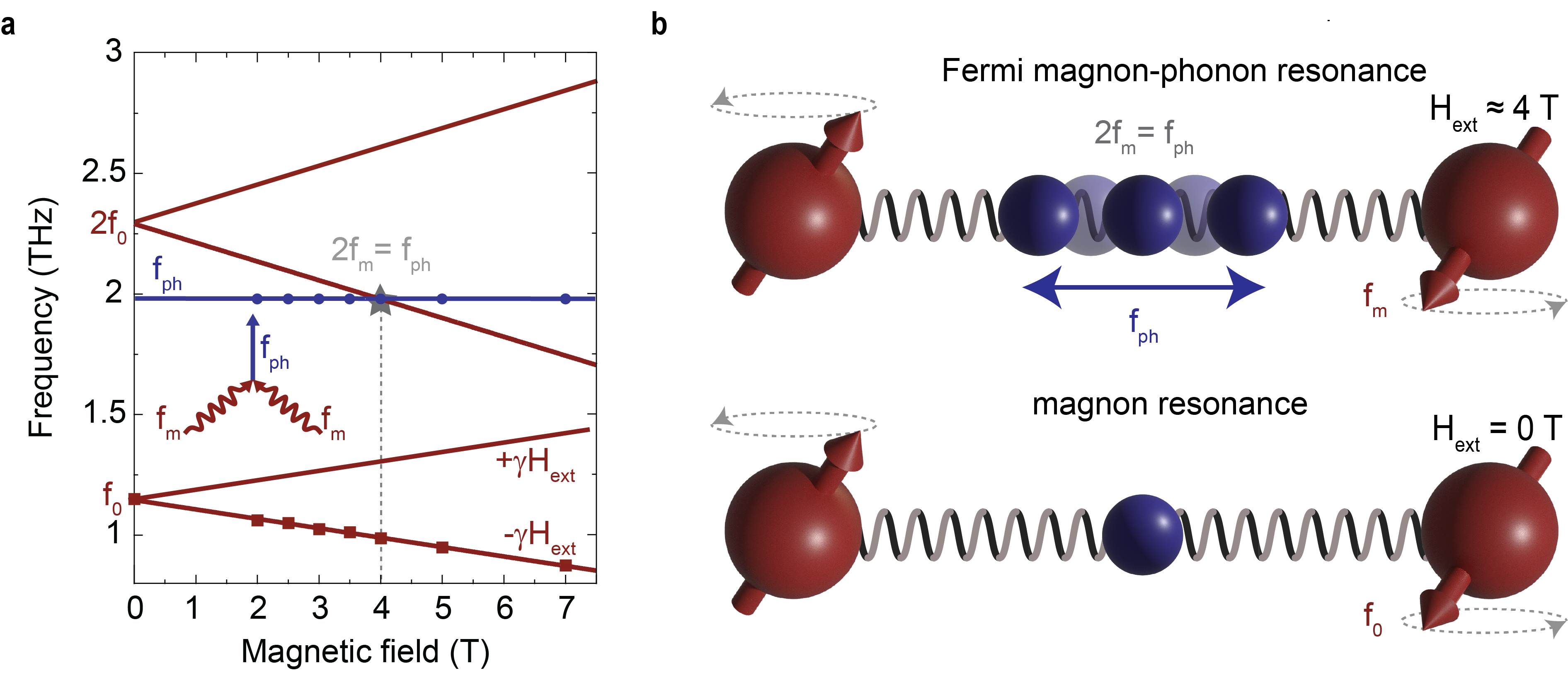} 
        \caption{\label{Fig1} The path to Fermi magnon-phonon resonance. (a) The frequency tuning by an external magnetic field applied along the antiferromagnetic easy axis of \CoF2. The frequency matching condition is marked by a grey star. The Feynman diagram illustrates the two magnon - one phonon conversion process. (b) Sketch of nonlinear magnon-phonon dynamics. For $2f\textsubscript{m}=f\textsubscript{ph}$, the Fermi magnon-phonon resonance condition is fulfilled and the the nonlinear energy transfer channel between a magnon and a phonon opens.}
    \end{figure}

In 1931~\cite{fermi_ueber_1931}, Enrico Fermi reported about the interaction of seemingly non-interacting vibrational modes in carbon dioxide  CO\textsubscript{2} molecules whose frequencies differ by a factor of two. In particular, it was argued that such an interaction explains a splitting present in the Raman spectrum in terms of a coupling between a single and an overtone vibrational mode of molecules. It was suggested that this resonance should be accompanied by a resonant energy transfer between the modes~\cite{katsnelson_resonance_1987,gornostyrev_phase_1995,gornostyrev_stochastic_1999,katsnelson_fermi-resonance-like_2002,katsnelson_nonperturbative_2004}. The phenomenon of nonlinear coupling, satisfying the conditions set forth by Fermi, was also observed in magnon dynamics~\cite{barsukov_giant_2019} resulting in shape deformations and broadening of the respective magnon spectrum, as well as in nano-mechanical systems showing coupled vibrational dynamics~\cite{shoshani_anomalous_2017}.
Interestingly, the recently demonstrated nonlinear coupling between spin and lattice dynamics~\cite{mashkovich_terahertz_2021} suggests that the Fermi resonance condition can be fulfilled in antiferromagnets as well. In zero applied magnetic field the double magnon frequency 2$f_0$ of antiferromagnetic \CoF2 at~6~K is higher then the frequency $f\textsubscript{ph}$ of the B$\textsubscript{1g}$ phonon ($2f_0>f\textsubscript{ph}$, see Fig.~\ref{Fig1}(a))~\cite{cottam_spin-phonon_2019}. 
Applying an external field along the magnetic easy-axis is known to shift the magnon frequency down, while the phonon frequency remains unchanged. Particularly, a field of 4 T is expected to be sufficient to reach the frequency matching condition $2f\textsubscript{m}=f\textsubscript{ph}$, where signatures of nonlinear coupling might be satisfied, see Fig.~\ref{Fig1}(a). 
We anticipate that in the vicinity of the resonance, the nonlinear energy flow between spins and lattice reaches a maximum. 
Thus, performing THz pump-infrared (IR) probe spectroscopy in combination with simulations, we reveal the corresponding fingerprints of coherent energy exchange driven by an impulsive THz stimulus resulting from the magnon-phonon dynamics in the vicinity of the Fermi-resonance (see Fig.~\ref{Fig1}(b)).
We expect that the impulsive excitation regime is characterized by its own benchmarks and opens new opportunities to tailor the excitation to selectively drive specific modes within the material.
\section{Results}\label{sec2} 
 \CoF2 belongs to the class of antiferromagnets with a rutile-type crystallographic lattice~\cite{borovik-romanov_piezomagnetism_1959}, described by the $P4_{2}/mnm$ space group. The multi-atomic primitive cell forms 15 optical phonon modes~\cite{balkanski_infrared_1966,kroumova_bilbao_2003}: $A_{1g} \oplus A_{2g} \oplus A_{2u} \oplus B_{1g} \oplus 2 B_{1u} \oplus B_{2g} \oplus 3 E_{u} \oplus E_{g}$.  The lowest-lying Raman-active phonon mode has B\textsubscript{1g} symmetry and is centered at a frequency of $f\textsubscript{ph}\approx1.96$ THz at T~=~6~K. It is worth noting that the frequency of this mode remains the same in external magnetic field up to 7 T.
 
The spins of Co\textsuperscript{2+} ions are aligned along the crystallographic c-axis below the N{\'e}el temperature of T\textsubscript{N}~=~39~K. In our experiment, we use a 500 um-thick single crystal \CoF2 plate cut perpendicular to the c-axis. If no magnetic field is applied, there is a doubly degenerate antiferromagnetic resonance mode at the frequency $f_0=1.14$ THz (at 6~K). Applying an external magnetic field, one breaks the degeneracy of the respective magnon mode. For instance, if a magnetic field is applied along the c-axis with its amplitude below the spin-flop field threshold~\cite{carretero-gonzalez_regular_1994}, the frequencies of two degenerate modes obey the relation $f\textsubscript{m}=f_0\pm \gamma H\textsubscript{ext}$, where $\gamma$ is the gyromagnetic ratio. 

\begin{figure}
        \centering
        \includegraphics[width=\columnwidth]{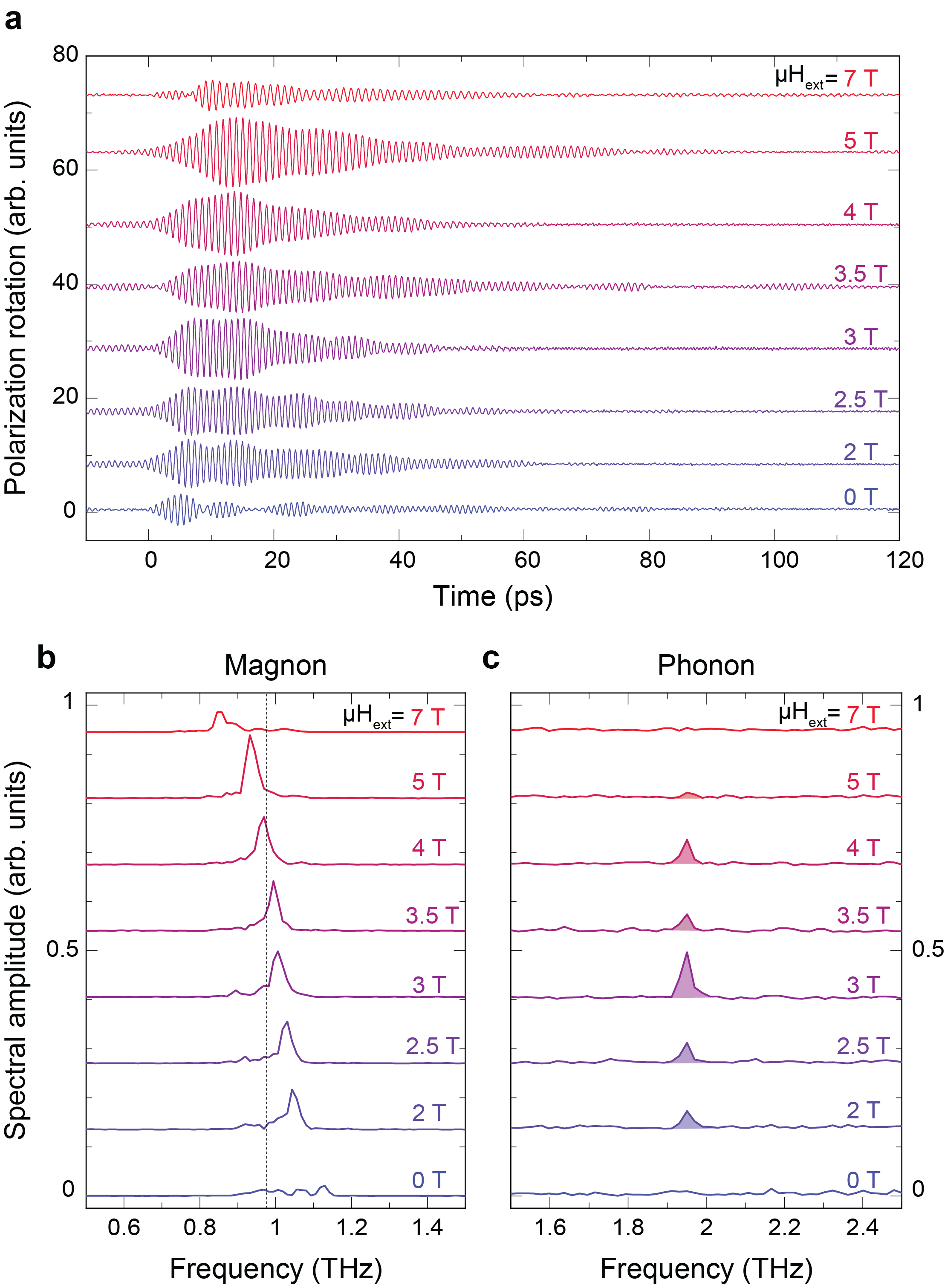} 
        \caption{\label{Fig2} (a) Time domain transients for the magnetic field applied along the c-axis and corresponding Fourier spectra in the vicinity of the magnon (b) and the phonon (c) resonance frequencies (T~=~6~K). The vertical line in panel (b) corresponds to half of the phonon frequency in panel (c). Two different time windows are used to highlight the magnon (-10 - 70 ps) and the phonon spectral peaks (70 - 120 ps). The phonon amplitude is scaled by a factor of 20 compared to the magnon amplitude. The plots for various external magnetic fields are offset by an equidistant constant. The shaded area correspond to the integrated region of interest for extraction of the spectral weight in Fig.~\ref{Fig3} (a).}
    \end{figure}

\begin{figure}
        \centering
        \includegraphics[width=\columnwidth]{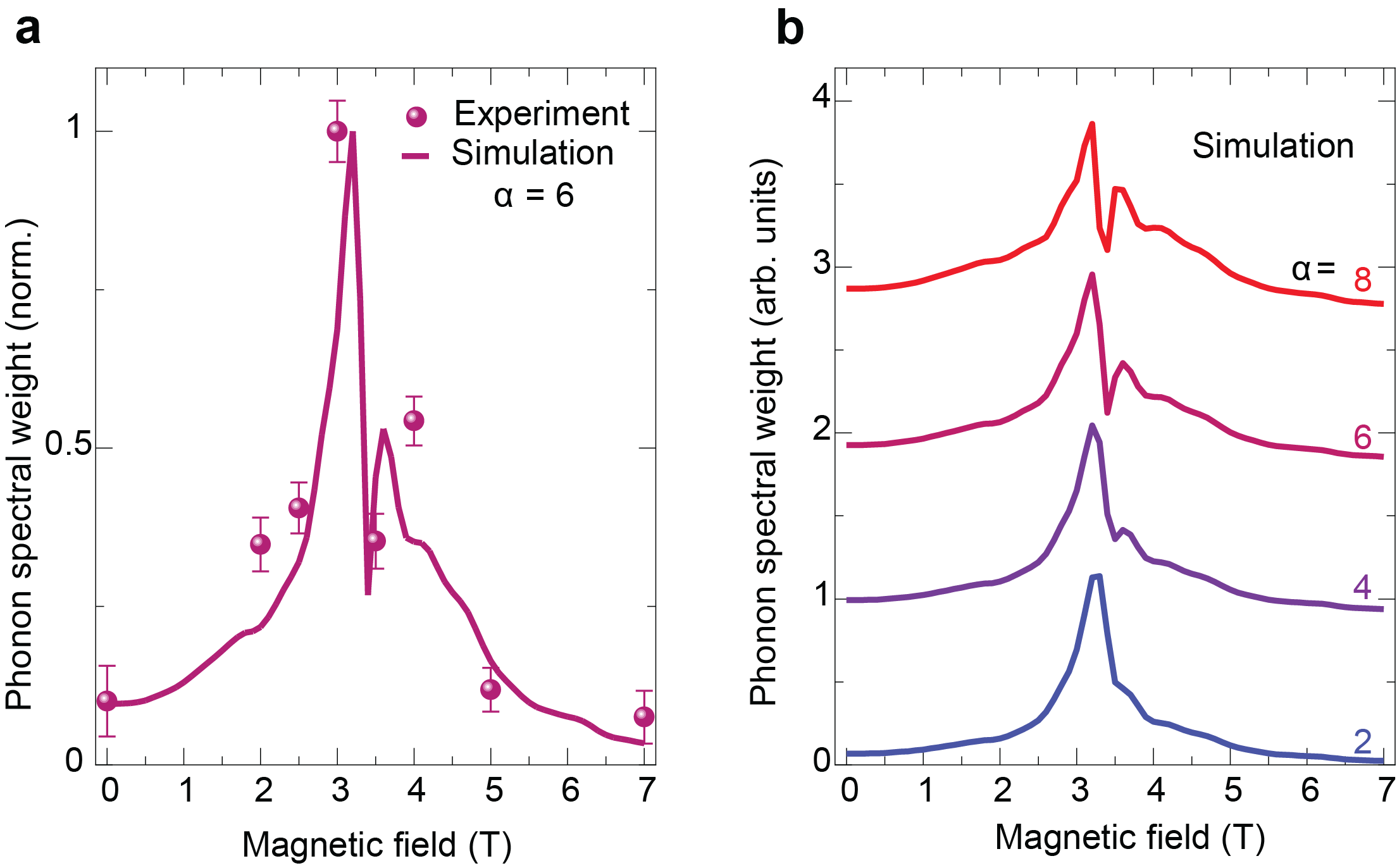}
        \caption{\label{Fig3}
        Fingerprints of the impulsive Fermi magnon-phonon resonance. (a) Phonon weight for the experiment (purple spheres) and simulation (purple line) for the nonlinear coupling constant $\alpha = 6$. The error bars are extracted as mean value of the FFT shown in Fig.~\ref{Fig2} (b) between 2 and 3 THz. (b) Simulated phonon spectral weight for various values of $\alpha$. Simulations are plotted according to equations in Appendix. 
        }
    \end{figure}

Using the intense, spectrally dense superradiant THz source TELBE located at Helmholtz-Zentrum Dresden-Rossendorf~\cite{green_high-field_2016} in combination with external magnetic fields $H\textsubscript{ext}$, we get the unique opportunity to pump the magnon selectively while controlling its center frequency $f\textsubscript{m}$ by tuning $H\textsubscript{ext}$ at our disposal. This configuration allows one to explore the spin-lattice interaction in the vicinity of the Fermi resonance $2f\textsubscript{m}=f\textsubscript{ph}$ by monitoring the phonon response. As it has been shown earlier~\cite{mashkovich_terahertz_2021}, as long as both the magnon and the phonon maintain their coherence, they will induce transient optical anisotropy in the originally isotropic (ab) plane of the antiferromagnet and thus modulate various components of the dielectric permittivity~\cite{metzger_effect_2022}, which we track by changes of the probe-polarization~\cite{mashkovich_terahertz_2021}. The experimental geometry and the THz pulse characteristics are provided in the Supplemental material.
   The THz-induced rotation of the probe-polarization measured in external magnetic fields up to 7 T (Fig.~\ref{Fig2}(a)) reveals the two distinct modes in the time domain data oscillating at frequencies $f\textsubscript{m}$ and $f\textsubscript{ph}$, respectively. The magnon amplitude dominates up to 70 ps whereas the 70 - 120 ps range is clearly dominated by the B\textsubscript{1g} phonon at 1.96 THz. In the Supplemental material we have zoomed into the region of 50 - 80  ps, where the magnon is fading, while the high frequency phonon oscillations are being revealed. 
    Performing a Fourier transformation (FFT) of the data presented in Fig.~\ref{Fig2} (a), we plot two spectral regions in Fig.~\ref{Fig2} (b,c) covering the magnon and the phonon modes, respectively. Evidently, the external magnetic field drastically changes the dynamics. Away from the Fermi-resonance at 0~T and 7~T the THz-pump pulse induced rotation shows the forced magnetic response closely following the magnetic field of the THz pulse (see Supplemental material) and no phonon induced dynamics is observed. Closer to the resonance for the in-between magnetic fields of 2 - 5 T, we observe the low energy magnon branch $f\textsubscript{m}$ with its frequency linearly decreasing with external magnetic field. It is remarkable that the phonon spectral amplitude shows a strong field-dependence, which does not have to correlate directly with the magnon spectral amplitude. Moreover, the phonon spectral amplitude does not exhibit a single but a double resonance located at 3 T and 4 T. Integrating the area under the phonon spectra for different external magnetic fields $H\textsubscript{ext}$ we retrieve the behaviour of the phonon resonance curve shown in Fig.~\ref{Fig3} (a). Contrary to the continuous wave regime where Fermi resonance implies a settled energy redistribution between modes of phononic or magnonic origin~\cite{barsukov_giant_2019,katsnelson_resonance_1987}, the impulsive regime is represented by a coherent energy exchange that modulates the mode amplitudes over time. Hence, in the scenario where the phonon lifetime surpasses that of the magnon (as is the case here), the energy of the phonon following the fading of the magnon will be determined by this intricate dynamics.

In order to obtain a better insight on the observed fingerprints of the magnon-phonon Fermi resonance, we undertook a simulation of the signatures of nonlinearly coupled dynamics in \CoF2. Conventionally, antiferromagnetic spins are described in terms of the N{\'e}el vector $\mathbf{L}=\mathbf{M}_{1}-\mathbf{M}_{2}$, where the net magnetic moments $\mathbf{M}_{1,2}$ are formed by Co\textsuperscript{2+} ions at the center and the corners of the unit cell, correspondingly~\cite{erickson_neutron_1953}. The movement of the \B1g phonon is characterized by the phonon coordinate $\theta\textsubscript{ph}$. The perturbations from the ground state are introduced as $\mathbf{L}(t)=(l_x(t),l_y(t),L_0)$. Here, L$_0$ describes the ground state N{\'e}el vector. The rule of Fermi resonance symmetry implies that the \B1g phonon symmetry ($x^2-y^2$) should follow the symmetry of the double magnon excitation.  Hence, the corresponding  nonlinear term can be introduced in the Lagrangian as $\Phi=-\alpha (l_x^2-l_y^2)\theta_{ph}$~\cite{breitenberger_elastic_1981}, where $\alpha$ represents the nonlinear coupling constant between the magnon and the phonon subsystems. We assume that the magnetic field of the THz-pulse $\bm{h}\textsubscript{THz}=(h_x,h_y,0)$ is polarized exclusively in the sample plane and solve the Lagrange-Euler equations, taking into accounting circularly polarized magnon states $l_{\pm}=l_x\pm i l_y$. The resulting coupled equations can be written as

\begin{multline}
        \frac{d^2l_+}{dt^2}+2\zeta_\textsubscript{m}\frac{dl_+}{dt}+\left(\omega_0^2-\gamma^2H_\textsubscript{ext}^2\right)l_++2\gamma i H_\textsubscript{ext}\frac{dl_+}{dt}=-2\alpha \theta_\textsubscript{ph}l_-+\gamma\frac{d}{dt}(h_y-i h_x),
        \label{eq:coupledosci1}
    \end{multline}
    \begin{multline}
        \frac{d^2l_-}{dt^2}+2\zeta_\textsubscript{m}\frac{dl_-}{dt}+\left(\omega_0^2-\gamma^2H_\textsubscript{ext}^2\right)l_--2\gamma i H_\textsubscript{ext}\frac{dl_-}{dt}=2\alpha \theta_\textsubscript{ph}l_++\gamma\frac{d}{dt}(h_y+i h_x) ,
        \label{eq:coupledosci2}
    \end{multline}
    \begin{equation}
        \frac{d^{2}\theta_{ph}}{dt^{2}}+2\zeta_\textsubscript{ph}\frac{d\theta_{ph}}{dt}+\omega_\textsubscript{ph}^{2}\theta_\textsubscript{ph} = -\alpha\left(l_+^2+l_-^2\right),
        \label{eq:coupledosci3}
     \end{equation}
\newline
     where $\omega_i=2\pi f_i$ and the Gilbert damping factors $\zeta_{i}$ with $i$ = "m" or $i$ = "ph" account for the magnon or the phonon subsystem, correspondingly. The second term on the right-hand side of Eqs.~\ref{eq:coupledosci1}-\ref{eq:coupledosci2} represents the linear Zeeman torque~\cite{satoh_spin_2010, baryakhtar_dynamics_1985}, while the nonlinear coupling can be introduced in Eqs.~\ref{eq:coupledosci1}-\ref{eq:coupledosci3} as the derivative of $\Phi$ on the corresponding order parameter. These terms represent the mutual nonlinear perturbation of the magnon (phonon) subsystem by the phonon (magnon) subsystem. The THz pulse waveform measured by electro-optical sampling (EOS) is introduced as the driving force in our simulation (see Supplemental material). 
            
Following a similar analysis as for the experimental data, we consider two time windows, where either the magnon or the phonon are dominating the dynamics. Performing the Fourier transform and extracting the phonon spectral weight (see the Supplemental material), we plot the phonon weight vs the externally applied magnetic field in Fig.~\ref{Fig3}(b). 
The THz field corresponds to the experiment and is of the order of 100~kV/cm. Accounting for g-factor of 2.6 \cite{strempfer_magnetic_2004} the gyromagnetic ratio is $\gamma$ = 36.4~GHz/Tesla. The Gilbert damping factors $\zeta\textsubscript{m}/2\pi$ = 10 GHz and $\zeta\textsubscript{ph}/2\pi$ = 5 GHz were estimated from fitting the magnon and the phonon experimental spectra (see Supplemental material).

Hence, the only unidentified variable $\alpha$ serves as the only fitting parameter.
Our simulations show that for values of $\alpha < 3$, the phonon amplitude exhibits only a single resonance peak. This demonstrates that the THz pulse energy is flowing into the magnon succeeded by a transfer into the phonon subsystem.
For $\alpha \geq 3$, a pronounced dip in the phonon spectral amplitude in the vicinity of the Fermi-resonance emerges, evidencing that nonlinear terms in Eqs. 1-2 are now affecting the dynamics.  
This in fact means that anharmonicty of the phonon starts to play a role and the energy is flowing back from the phonon into the magnon subsystem.
However, at higher values of $\alpha>>8$, the nonlinearity becomes strong and perturbs the form of the phonon spectrum.
\section{Discussion}\label{sec4}
Despite the fact that nonlinear phononics is a relatively young field, Fermi resonance of the lattice has been studied theoretically earlier~\cite{katsnelson_resonance_1987,gornostyrev_phase_1995,gornostyrev_stochastic_1999, katsnelson_fermi-resonance-like_2002,katsnelson_nonperturbative_2004}. In particular, it was shown that the resonances can result either in a splitting or broadening of phonon lines in the vibrational spectra of the lattice. Moreover, according to Ref.~\cite{barsukov_giant_2019}, in the vicinity of the resonance, nonlinear damping plays a significant role and in principle can modify a single bell-shaped resonance phonon curve.  

A double or Fermi resonance between otherwise non-interacting phononic modes can be practically regarded as evidence of energy exchange between these modes. In strong contrast to the previous theoretical studies focused on incoherent lattice dynamics and thus revealing stochastic acts of energy exchange between the modes~\cite{gornostyrev_stochastic_1999, barsukov_giant_2019}, our experiments reveal the manifestation of the Fermi resonances for the case of coherent impulsive dynamics. Thus, using the methods of coherent control one can, in general, steer the energy flow between spins and lattice on demand. Moreover, deriving the magnon-phonon quasi-particles conservation law analogous to the Manley-Rowe relations for second harmonic generation~\cite{manley_general_1956} is an intriguing aspect.
\section{Conclusion}\label{sec5}
In summary, we reported about a new regime of magnon-phonon dynamics which facilitates an efficient energy exchange between the magnons and phonons fulfilling the conditions of Fermi resonance. We revealed the fingerprints of Fermi magnon-phonon resonance and showed, in particular, a distinct double-resonance feature of the phonon amplitude upon sweeping the frequency of the magnon resonance by applying an external magnetic field indicating the effect of a mutual, anharmonic energy transfer. We believe that our finding is not only a new phenomenon in nonlinear phononics, but also an important milestone in the broad fields of magnonics and spintronics, where coherent energy control plays the central role. 
\backmatter
\bmhead{Acknowledgments}
This work was supported by the Deutsche Forschungsgemeinschaft (DFG, German Research Foundation) - Project number 277146847 - CRC 1238, de Nederlandse Organisatie voor Wetenschappelijk Onderzoek (NWO), the European Union’s Horizon 2020 research and innovation program under the Marie Skłodowska-Curie grant agreement No 861300 (COMRAD), the European Research Council ERC Grant Agreement No.101054664 273 (SPARTACUS), and RSF (grant No. 21-42-00035). MIK and AVK acknowledge the research program "Materials for the Quantum Age" (QuMat) for financial support. This program (registration number 024.005.006) is part of the Gravitation program financed by the Dutch Ministry of Education, Culture and Science (OCW). RMD acknowledges support of RSF (Grant No. 22-72-00025) and the Ministry of Science and Higher Education of the Russian Federation (FSWR-2021-011). 
Parts of this research were carried out at ELBE at the Helmholtz-Zentrum Dresden - Rossendorf e. V., a member of the Helmholtz Association. The TELBE measurements were conduced February 11-15, 2022 proposal number 21202639-ST and October 23-25, 2021 proposal number 21102489-ST. We thank Dr. Clément Faugeras for discussions. The authors declare that this work has been published as a result of peer-to-peer scientific collaboration between researchers. The provided affiliations represent the actual addresses of the authors in agreement with their digital identifier (ORCID) and cannot be considered as a formal collaboration between the aforementioned institutions. 

\bibliography{FermiSpinLatticePendulumV16BibTex.bib}
\end{document}